# Near-Room-Temperature Ferromagnetic Behavior of Single-Atom-Thick 2D Iron in Nanolaminated Ternary MAX Phases

*Youbing Li, Jinghua Liang, Haoming Ding, Jun Lu, Xulin Mu, Pengfei Yan, Xiao Zhang, Ke Chen, Mian Li, Per O. Å. Persson, Lars Hultman, Per Eklund, Shiyu Du, Hongxin Yang,\* Zhifang Chai, Qing Huang\**

Dr. Y. B. Li, H. M. Ding, X. Zhang, Dr. K. Chen, Dr. M. Li, Prof. S. Y. Du, Prof. Z. F. Chai, Prof. Q. Huang
Engineering Laboratory of Advanced Energy Materials
Ningbo Institute of Industrial Technology
Chinese Academy of Sciences
Ningbo, Zhejiang 315201, China
E-mail: huangqing@nimte.ac.cn
Dr. Y. B. Li, H. M. Ding, X. Zhang, Dr. K. Chen, Dr. M. Li, Prof. S. Y. Du,
Prof. Z. F. Chai, Prof. Q. Huang
Qianwan Institute of CNiTECH
Zhongchuangyi Road, Hangzhou bay District
Ningbo, Zhejiang, 315336, China
Dr. J. H. Liang, Prof. H. X. Yang
Center of Materials Science and Optoelectronics Engineering
University of Chinese Academy of Sciences
Beijing 100049, China
E-mail: yanghongxin@nimte.ac.cn
Prof. J. Lu, Prof. P. O. Å. Persson, Prof. L. Hultman, Prof. P. Eklund
Department of Physics, Chemistry, and Biology (IFM)
Linköping University
Linköping 58183, Sweden
X. L. Mu, Prof. P. F. Yan
Beijing Key Laboratory of Microstructure and Properties of Solids
Beijing University of Technology
Beijing 100124, China
Y. B. Li and J. H. Liang contributed equally to this work

**Abstract:** Two dimensional (2D) ferromagnetic materials have attracted much attention in the fields of condensed matter physics and materials science, but their synthesis is still a challenge given their limitations on structural stability and susceptibility to oxidization. MAX phases nanolaminated ternary carbides or nitrides possess a unique crystal structure in which single-atom-thick "A" sublayers are interleaved by two dimensional "MX" slabs, providing nanostructured templates for designing 2D ferromagnetic materials if the non-magnetic "A" sublayers can be substituted replaced by magnetic elements. Here, we report three new ternary magnetic MAX phases ($Ta_2FeC$, $Ti_2FeN$ and $Nb_2FeC$) with "A" sublayers of single-atom-thick 2D iron through an isomorphous replacement reaction of MAX precursors ($Ta_2AlC$, $Ti_2AlN$ and $Nb_2AlC$) with a Lewis acid salts ($FeCl_2$). All these MAX phases exhibit ferromagnetic (FM) behavior. The Curie temperature ($T_c$) of $Ta_2FeC$ and $Nb_2FeC$ MAX phase are 281 K and 291 K, respectively, i.e. close to room temperature. The saturation magnetization of these ternary magnetic MAX phases is almost two orders of magnitude higher than that of $V_2(Sn,Fe)C$ MAX phase whose A-site is partial substituted by Fe. Theoretical calculations on magnetic orderings of spin moments of Fe atoms in these nanolaminated magnetic MAX phases reveal that the magnetism can be mainly ascribed to intralayer exchange interaction of the 2D Fe atomic layers. Owning to the richness in composition of MAX phases, there is a large compositional space for constructing functional single-atom-thick 2D layers in materials using these nanolaminated templates.

**Keywords:** MAX phases, isomorphous replacement, magnetic elements, 2D ferromagnetic materials

## 1. Introduction

Since the synthesis of two-dimensional (2D) ferromagnetic (FM) monolayers in experiments, *e.g.*, CrI$_3$[1] and Cr$_2$Ge$_2$Te$_6$,[2] 2D FM materials have attracted much attention in the field of condensed matter physics and materials science, as they provide a good opportunity to explore the fundamental physical properties of magnetic materials with quantum confinement effect, and are promising for magnetic applications, e.g., spintronics and magnetic storage devices. One of the challenges in these materials is how to construct a specific crystal structure that realizes spin-electron interaction in a confined 2D space. The 'MAX phases' are a suitable class of model systems for realizing this general concept. These phases are a family of inherently nanolaminated ternary carbides and/or nitrides with a hexagonal crystal structure (space group P6$_3$/mmc), and a formula of M$_{n+1}$AX$_n$ (where M is an early transition metal, A is mainly an element from groups 13-16, and X is carbon or/and nitrogen, n=1-3).[3-4] One of the most fascinating features of the MAX phases, is that single-atom-thick sheets of the A element are repetitively stacked with edge-sharing [M$_6$X] octahedra slabs. Due to the weak interaction between A atoms and nearest M atoms, MAX phases could be an ideal template to construct 2D ferromagnetic monolayers if the A layers are composed of magnetic elements. However, although nearly 90 ternary MAX phases have been reported so far,[5] no ternary MAX phases have ever been achieved with exclusively magnetic elements (such as Fe, Co, or Ni) occupying the A-sites.

Recently, we reported a series of nanolaminated V$_2$($A_x$Sn$_{1-x}$)C (*A*=Fe, Co, Ni and Mn, and combinations thereof, with x≈1/3) MAX phases by an competing-alloy-guided reaction, in which Fe/Co/Ni/Mn atoms were partially incorporated into the A layers of MAX phases through competitively alloying with tin other than with transition metal vanadium.[6] However, simulated V-A-C (A=Fe) ternary phase diagrams predicted that ternary V$_2$AC phases (A=Fe) are not thermodynamically competitive to other alloys and binary carbides [6], precluding their formation through traditional metallurgical synthesis. Recently, in order to overcome the interference of competing phases on the formation of targeting MAX phases, an isomorphous replacement approach was developed. This replacement strategy takes advantage of chemically stable $M_{n+1}X_n$ slabs, with a full replacement reaction of the A atoms between these slabs in nanolaminated MAX phases by annealing in contact with a metal reservoir [7-10], or a Lewis acid molten salt [11-13] to achieve atomic-layer-exchange in a confined single-atom-thick 2D space. Since H. Notwotny *et al*. reported the first MAX phase in the 1960's[14-15], this isomorphous replacement strategy has resulted in groundbreaking new ternary MAX phases (such as $Ti_3AuC_2$[7], $Ti_3IrC_2$[7], $Mo_2AuC$[9], $Ti_3ZnC_2$[11], $Ti_2ZnC$[11], $V_2ZnC$[11], $Ti_2ZnN$[11] and $Nb_2CuC$[13]). Moreover, since the element replacement totally happened in a single-atom-thick spacing ("A" layer), such material fabrication approach provides a new synthesis solution to construct 2D atom arrangement that is hardly realized through neither "bottom-up" nor "top-down" methodologies.

In this communication, we use isomorphous replacement to synthesize three new ternary MAX phases $Ta_2FeC$, $Ti_2FeN$, and $Nb_2FeC$, with Fe atoms exclusively occupying the two-dimensional "A" layers in the MAX phases. Experimental results clearly shows that 2D arrangement of Fe atoms in these nanolaminated MAX phases exhibit ferromagnetic properties with Curie temperatures ($T$c) even close to room temperature. Theoretical calculations reveal that the magnetism is mainly attributable to intra-layer exchange interaction of constructed 2D iron atomic layers.

## 2. Results and Discussion
### 2.1 Preparation and Characterization of MAX phases
The detailed experimental conditions can be found in Supporting Information (SI), Section 1 and Table S1. The XRD patterns of as-prepared novel MAX phases by an isomorphous replacement approach are shown in Figure 1a. Substitution of Fe atoms for Al atoms in final MAX phases makes no change in crystal structure what is evidenced by almost identical X-ray diffraction patterns but have influence on element-depending capability of electron scattering at specific indices of crystal faces (such as {000$l$} in MAX phases). For Ta$_2$FeC, it can be found that the diffraction peak positions are close to that of original Ta$_2$AlC MAX phase (Figure 1a, orange line), but the intensity of (002) and (004) planes diminished below the detection limit. A Rietveld refinement analysis was performed on the obtained XRD pattern of the final product and a reliability factor $R_{wp}$ as low as 8.06 % is achieved (Figure 1a), corroborating an accurate occupancy of Fe atoms on original Al atom positions. The lattice parameters of Ta$_2$AlC are $a$ = 3.075 Å and $c$ = 13.83 Å,[14] whereas the lattice parameters determined from the Rietveld refinement of Ta$_2$FeC are $a$ = 3.103 Å and $c$ = 13.70 Å. The atom positions of Ta$_2$FeC are shown in Table S2 (Supporting Information). Morphology of Ta$_2$FeC particles (Figure 1b) display the typical layered structure of the Ta$_2$AlC (Figure S1, Supporting Information). Energy-dispersive X-ray spectroscopy (EDS) results indicate that the products contain only Ta, Fe and C element, without Al element (Figure S2, Supporting Information). EDS element mapping of a typical particle (Figure S3, Supporting Information) show uniform element distribution of Ta, Fe, and C in a Ta$_2$AlC-converted product, and semi-quantitative analysis of atomic ratio of Ta:Fe:C gives close to 2:1:1, which is the retained stoichiometry of Ta$_2$AlC precursor, indicating the full-replacement of Al by Fe in the resultant product.

The atomic structure and chemical analysis of the as-synthesized Ta$_2$FeC were further identified by a high-resolution high-angle annular dark field scanning transmission electron microscopy (HAADF-STEM) together with lattice resolved energy dispersive X-ray (EDS) spectroscopy. Viewed along the [11$\bar{2}$0] orientation, the bright Ta columns exhibit the MAX phase archetype zig-zag pattern along the c axis, separated by single atomic thick layer of the A elements (Table S2, Supporting Information). Since the brightness of the atom is directly related to its mass (intensity~$Z^2$), the A atoms (both Al and Fe) are barely distinguishable between the Ta$_2$C slabs. However, lattice-resolved EDS elemental mapping (of Ta-$L\alpha$ and Fe-$K\alpha$ photons) and an integrated line profile through verified that the A sites are occupied by Fe (Figure 1e).

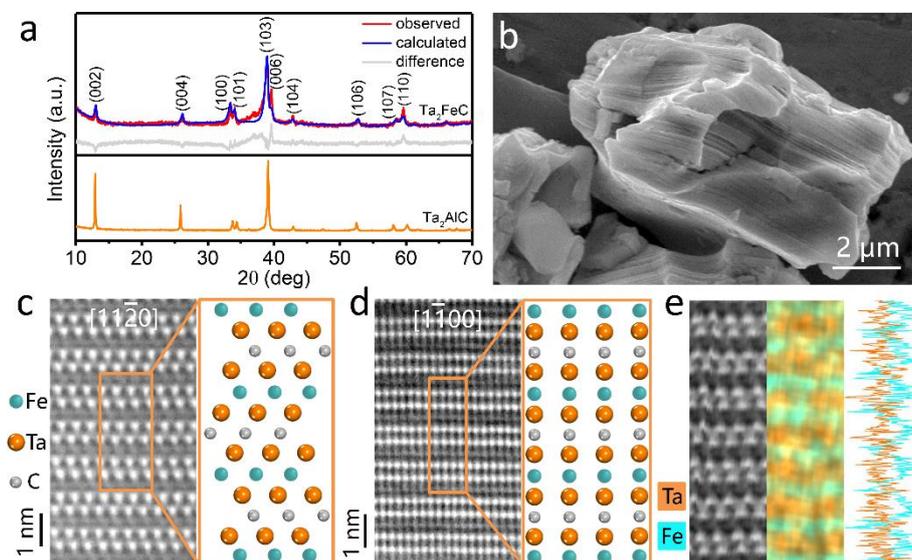

**Figure 1.** Phase compositions and microstructure of new magnetic MAX phases. a) XRD pattern of Ta$_2$AlC and Rietveld refinement of XRD of Ta$_2$FeC. b) SEM image of Ta$_2$FeC. Aberration corrected (HR)-STEM imaging and lattice resolved EDS elemental mapping of the MAX structure. High-resolution (HR)-STEM images of Ta$_2$FeC showing atomic columns along (c) [11$\bar{2}$0] and (d) [1$\bar{1}$00] direction, respectively. e) STEM-EDS mapping and integrated line profile of Ta-$L_\alpha$ (red) and Fe-$K_\alpha$ (green), respectively, in [11$\bar{2}$0] projection.

The rich variation of MAX phases can be attributed to the flexible choice of M elements from early transition metals (*e.g.*, Ti and Nb) and of X elements from carbon or nitrogen. Herein, Ti$_2$AlN and Nb$_2$AlC precursors were selected to further explore the generality of our isomorphous replacement approach, and both were finally converted into Ti$_2$FeN and Nb$_2$FeC. Figure 2 shows the phase identification of products derived from M$_2$AX (Ti$_2$AlN and Nb$_2$AlC) precursors. The XRD patterns in Figure 2a-2b exhibit a similar variation to that of the Ta$_2$AlC-FeCl$_2$ system, with the relative intensity of the (000*l*) peaks changed. Rietveld refinement of XRD patterns of the Ti$_2$FeN (Figure 2a) and Nb$_2$FeC (Figure 2b) MAX phases, yields a reliability factor $R_{wp}$ of 5.70% and 7.07%, respectively. The calculated lattice parameters are *a*

= 3.0066 Å and $c$ = 13.6307 Å for $Ti_2FeN$, $a$ = 3.1301 Å and $c$ = 13.7415 Å for $Nb_2FeC$. The microstructures of $Ti_2FeN$ (Figure S4b, Supporting Information) and $Nb_2FeC$ (Figure S5b, Supporting Information) are consistent with the precursor MAX phases $Ti_2AlN$ (Figure S4a, Supporting Information) and $Nb_2AlC$ (Figure S5a, Supporting Information), and corresponding EDS results (Figure S6 and Figure S7, Supporting Information) show the absence of Al element in both products. The STEM images (Figure S8 and Figure S9, Supporting Information) and the EDS mapping (Figure 2c-2d) further confirmed the success in formation of $M_2FeX$ phases ($Ti_2FeN$ and $Nb_2FeC$) after the isomorphous replacement reactions. Corresponding details can be found in supporting information (Figure S4-S9 and Table S3-S4).

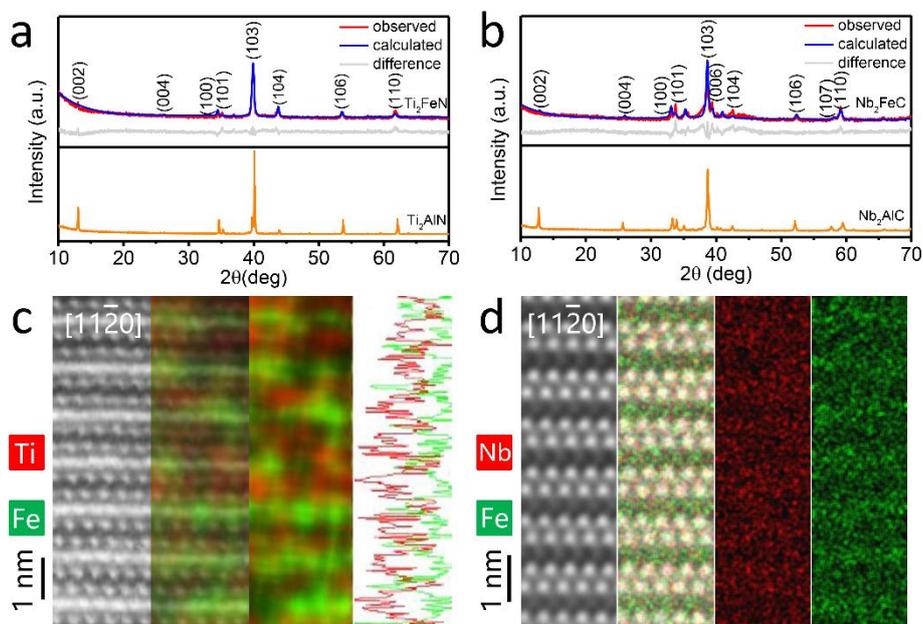

**Figure 2.** Phase composition and microstructure of new magnetic MAX phases. a) XRD pattern of $Ti_2AlN$ and Rietveld refinement of XRD of $Ti_2FeN$. b) XRD pattern of $Nb_2AlC$ and Rietveld refinement of XRD of $Nb_2FeC$. Aberration corrected (HR)-STEM imaging and lattice resolved EDS elemental mapping of the MAX structure. c) STEM-EDS mapping and integrated line profile of Ti-$K_\alpha$ (red) and Fe-$K_\alpha$ (green), respectively, in the [11$\bar{2}$0] projection. d) STEM-EDS mapping of Nb-$L_\alpha$ (red) and Fe-$K_\alpha$ (green), respectively, in the [11$\bar{2}$0] projection.

## 2.2. Magnetic Properties of MAX phases

In order to understand the physical behavior of 2D Fe in these nanolaminated MAX phases, the magnetic properties of $Ta_2FeC$, $Ti_2FeN$, and $Nb_2FeC$ are shown in Figure 3. All tested samples were soaked in HCl solution for 2h to ensure that there is no presence of residual iron or iron-containing alloy impurities before magnetic property measurement (Take $Ti_2FeN$ as example, Figure S10, Supporting Information). Temperature dependent magnetization M(T) curves under zero-field-cooled (ZFC) mode at a magnetic field of 1000 Oe are shown in Figure 3a, marked black ($Ta_2FeC$), blue ($Ti_2FeN$), and red ($Nb_2FeC$). In Figure 3a, a decrease of magnetization with increasing temperature is observed, which indicate that $Ta_2FeC$, $Ti_2FeN$, and $Nb_2FeC$ MAX phases have a typical ferromagnetic-paramagnetic transition behavior. Here, the Curie temperature ($T$c) of $Ta_2FeC$ and $Nb_2FeC$ was 291 K and 281 K, respectively. These values are close to room temperature, while the $T$c of $Ti_2FeN$ was only 208 K. In contrast, $T_c$ of Fe bulk is beyond 1000 K, which excludes the possibility of the measured $T_c$ being artifactual from residual Fe. These are very high $T$c values among reported magnetic MAX phases. The $T$c of magnetic MAX phases with Mn or Fe in M-sites are relatively low in previous reportes, $e.g.$, $Mn_2GaC$[16] ($T$c < 230 K), $(Cr_{0.85}Mn_{0.15})GeC$[17] ($T$c = 205 K), $(Cr_{1.95}Fe_{0.05})GeC$[18] ($T$c = 250 K), $(Cr,Fe)_2AlC$[19] ($T$c = 40 K), and $(Cr_{0.5}Mn_{0.5})_2AuC$[20] ($T$c = 100 K). Moreover, the splitting of the ZFC and FC curves for these three MAX phases is also visible (Figure S11, Supporting Information), which is a characteristic feature of spin glass. The shape of M(T) curve (Figure S10, Supporting Information) may be the result of weak antiferromagnetic (AFM) interaction, which competes with ferromagnetic (FM) interaction, leading to non-collinear spin and glassy behavior.[21] This may again explain the glassy behavior of magnetic MAX phases seen in ZFC-FC measurements.

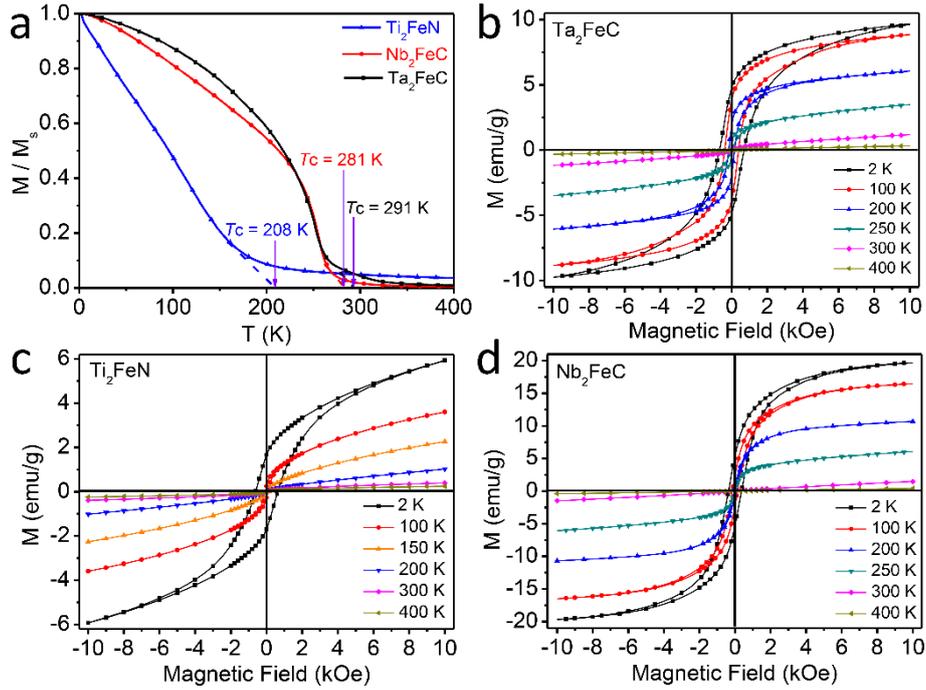

**Figure 3.** Magnetic response measured with a SQUID sample magnetometer for different temperatures. a) Zero-field-cooling (filled circles) curves for the Ta$_2$FeC (black lines), Ti$_2$FeN (blue lines), and Nb$_2$FeC (red lines) at 1000 Oe with temperature in the range of 2-400 K. Magnetic hysteresis loops of Ta$_2$FeC (b), Ti$_2$FeN (c) and Nb$_2$FeC (d) at different temperatures in the range from -10 kOe to 10 kOe.

The magnetic hysteresis loops of Ta$_2$FeC, Ti$_2$FeN, and Nb$_2$FeC dependent on coercive force ($H_c$) in the range of -10 kOe -10 kOe and temperature in the range of 2-400 K are shown in Figure 3b-3d, respectively. With increasing magnetic field the magnetization continues to rise and does not reach saturation even at 10000 Oe. A high saturation field can be a signature of a large random anisotropy because of variations in the alignment of microcrystalline grains.[22] The residual magnetization ($M_r$) and maximum saturation magnetization ($M_s$) are both decrease gradually with increasing temperature (Table S5-S7, Supporting Information). The "S-shape" behavior and a reduced residual magnetization of those MAX phases are similar to other magnetic MAX phases.[6,23,24] This indicates a non-collinear spin configuration in small external fields which most probably attributes to both competing FM and AFM interactions in ferromagnetic M-X-M networks separated by the A element and the Zeeman energy in the applied magnetic field favoring FM alignment.[25,26] Noteworthily, magnetic MAX phases, such as (Cr,Mn)$_2$GeC[23] and Mn$_2$GaC[16], did not exhibit magnetic response at temperature above 230 K, while the magnetic response of present Nb$_2$FeC and Ta$_2$FeC remains above 300 K. In the following discussions, we will see that the high Curie temperature of Nb$_2$FeC and Ta$_2$FeC mainly comes from strong ferromagnetic exchange coupling in the 2D Fe atom layer. Moreover, the M$_s$ (2 K, 10 kOe) of Nb$_2$FeC, Ta$_2$FeC, and Ti$_2$FeN are 19.69, 9.63, and 5.93 emu/g, which amount to 0.89, 0.74, and 0.18 $\mu_B$/Fe (Table S5-S8, Supporting Information), respectively. Compared to the partial substitution of magnetic element in A-site of V$_2$(Sn,*A*)C MAX phases,[6] the saturation magnetization of these ternary magnetic MAX phases are near 2 orders of magnitude higher, and also higher than most solid solution magnetic MAX phases such as (Cr$_{0.5}$Mn$_{0.5}$)$_2$AuC,[20] (Cr$_{0.5}$Mn$_{0.5}$)$_2$GaC,[21] (Mo$_{0.5}$Mn$_{0.5}$)$_2$GaC.[25] It may be that the overlap between electron clouds of M and A atoms is limited compared to that between M and X atoms, which should aid in preserving the magnetic properties. Thus, this approach provides a route for altering the magnetic properties of MAX phases by changing the chemical composition and component in M atomic layers.

**2.3 First-Principles Calculations**

We perform first-principles calculations with the Vienna ab initio simulation package (VASP)[27] in order to understand the magnetic properties of M$_2$AX (A=Fe; M=Ta, Nb, Ti; X=C, N) compounds. The exchange correlation functional in form of PBEsol[28] is used. For Ti-3*d* orbitals, an effective Hubbard U = 3 eV is added. The plane-wave cutoff energy of 520 eV is used in all calculations. To determine the ground state of the compounds, we have considered three magnetic orderings of spin moments of Fe atoms, *e.g.*, the ferromagnetic (FM), intralayer (AFM1) and interlayer (AFM2) antiferromagnetic states, using a 2 × 1 × 1 supercell (see Figure 4a). The full Brillouin zone is sampled using a Γ-centered 8 × 16 × 4 Monkhorst-Pack k-point grid.[29] All structures are fully relaxed until residual forces are smaller than 0.001 eV/Å. The spin Hamiltonian of the MAX phase can be described as$=-\sum_{<i,j>}J_1 S_i \cdot S_j - \sum_{<i,k>}J_2 S_i \cdot S_k$, where $S_i$ is the unit vector along the direction of magnetic moment of Fe atom. $J_1$ and $J_2$ denote the intralayer and interlayer exchange interaction parameters, respectively, as indicated in Figure 4a, and can be extracted by energy mapping analysis based on the considered magnetic configurations. Table S8 (Supporting Information) presents the calculated total energy given with respect to the ferromagnetic state and magnetic parameters of the $M_2AX$ compounds. Consistent with the above experiments, we find that $Ta_2FeC$, $Ti_2FeN$ and $Nb_2FeC$ all host ferromagnetic ground state, which is consistent with the experimental results, whereas $Ta_2FeN$, $Nb_2FeN$ and $Ti_2FeC$ are antiferromagnetic (Table S8, Supporting Information). We note that in all calculated MAX phases, the exchange interaction is dominated by the intralayer interaction since $J_1$ is always much larger than $J_2$, which implies that the magnetism of the MAX phases is confined within each A-site (Fe) layer. With the calculated exchange interaction parameters $J_1$ and $J_2$, we can determine the Curie temperature of ferromagnetic MAX phases via Monte Carlo simulations. Importantly, we find that since the intralayer exchange coupling $J_1$ of both $Ta_2FeN$ and $Nb_2FeN$ are almost twice as large as that of $Ti_2FeN$ (Table S8, Supporting Information), the calculated Curie temperatures (see Figure 4b for MC simulations and Table S8) of $Ta_2FeC$ (240 K) and $Nb_2FeC$ (210 K) are much higher than $Ti_2FeN$ (130 K), this tendency of $T_c$ ($Ta_2FeC > Nb_2FeC > Ti_2FeN$) is also in good agreement with the experimental results. To further analyze the underlying magnetic mechanism, we plot the projected density of states (DOS) of the ferromagnetic $Ta_2FeC$, $Ti_2FeN$ and $Nb_2FeC$ phases in Figure 4c. From the plotted spin polarized Fe-3$d$ states (red lines in Figure 4c), in which the majority spin channel is almost completely filled while the minority spin channel is only partially filled, it corroborates that the magnetism of the MAX phases is mainly attributable to the A-site Fe atoms. Moreover, the Fe-3$d$ states of $Nb_2FeC$, $Ta_2FeC$ and $Ti_2FeN$ have an exchange splitting $\Delta E_{ex}$ of 0.94 meV, 0.66 meV and 0.52 meV, respectively. Correspondingly, the calculated magnetic moments of Fe atoms decrease in the order of $\mu_{Fe}(Nb_2FeC) > \mu_{Fe}(Ta_2FeC) > \mu_{Fe}(Ti_2FeN)$ (Table S8, Supporting information) that is in line with our measured saturation magnetizations. From these calculations, we can conclude that the magnetism of the MAX phases can be ascribed to 2D single-atom-thick Fe layers, and the high Curie temperatures of Ta$_2$FeC and Nb$_2$FeC originate from their strong intralayer exchanging coupling. In contrast, the magnetic properties of the hypothetical Fe$_2$AlC and Fe$_2$AlN with Fe atoms in the M sites should be antiferromagnetic (Figure S12, Supporting Information), indicating the importance of coordination environment of Fe atom on their magnetics.

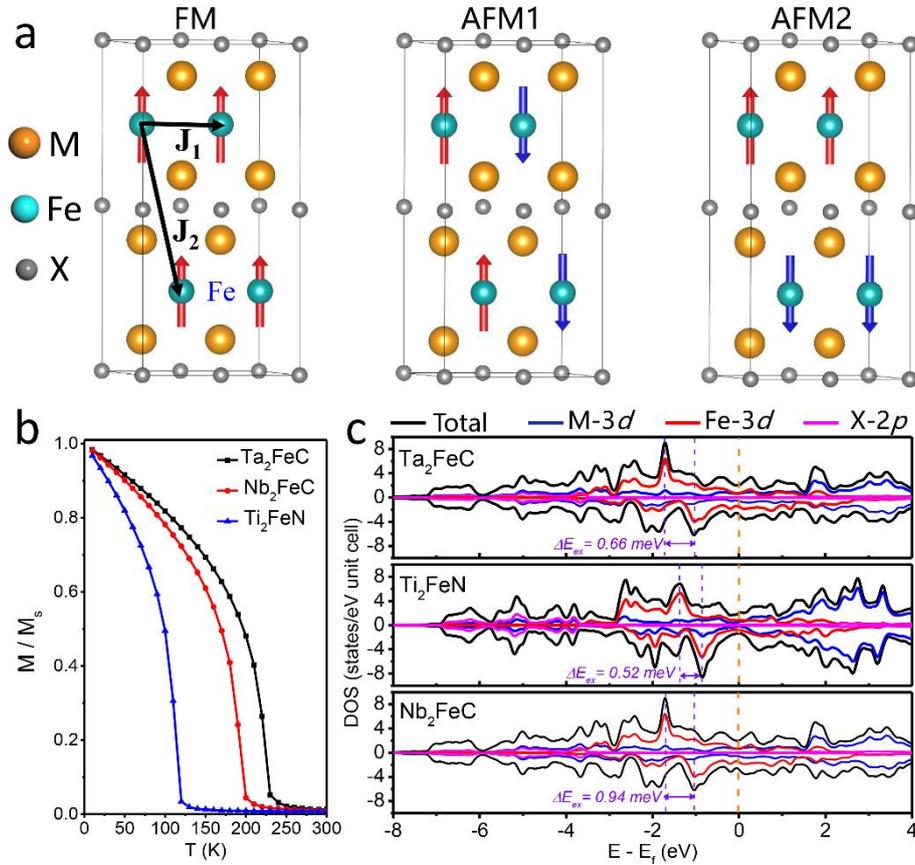

**Figure 4.** a) Considered ferromagnetic (FM), intralayer (AFM1) and interlayer (AFM2) antiferromagnetic states for MAX phases. b) MC simulations of normalized magnetization as a function of temperature and (c) the calculated projected DOS for Ta$_2$FeC, Ti$_2$FeN and Nb$_2$FeC, and the exchange splitting $\Delta E_{ex}$ of Fe-3$d$ states indicated by the purple dashed lines.

**Conclusions**

In conclusion, three new MAX phases Ta$_2$FeC, Ti$_2$FeN, and Nb$_2$FeC have been synthesized by isomorphous replacement of the A-element (Al) by Fe, which also realizes the formation of 2D single-atom-thick Fe layer in these ternary nanolaminated MAX phases. Ta$_2$FeC and Nb$_2$FeC MAX phases display characteristic ferromagnetic behavior and have Curie transition temperature close to room temperature ($T$c of 281 K and 291 K, respectively). Theoretical simulation indicates that the magnetism of the MAX phases is mainly ascribed to the strong intralayer exchange interaction of the 2D Fe single-atom-thick layers. Taking advantage of the naturally nanolaminated structure, high stability, highly anisotropic transport properties, and composition adjustability (selection of M, A, or X element and their alloying), MAX phases provide powerful templates for designing 2D atom arrangement and exploration of their electronic and spintronic properties via isomorphous replacement on A-layer.

**Supporting Information**

Supporting Information is available from the Wiley Online Library or from the author.

**Acknowledgements**

This study was supported financially by the National Natural Science Foundation of China (grant nos. 21671195, 91426304 and 11874059), International Partnership Program of Chinese Academy of Sciences (Grant No. 2019VEB0008), Leading Innovative and Entrepreneur Team Introduction Program of Zhejiang (Grant No. 2019R01003), Ningbo Top-talent Team Program, Ningbo Municipal Bureau of Science and Technology (Grant No. 2018A610005), Key Research Program of Frontier Sciences, CAS, Grant NO. ZDBS-LY-7021, Beijing National Laboratory for Condensed Matter Physics and Natural Science Foundation of Zhejiang Province (Grant No. LR19A040002). We acknowledge support from the Swedish Government Strategic Research Area in Materials Science on Functional Materials at Linköping University (Faculty Grant SFO-Mat-LiU No. 2009 00971). The Knut and Alice Wallenberg Foundation is acknowledged for support of the electron microscopy laboratory in Linköping (Grant KAW 2008.0058), a project grant (2020.0033) an Academy Fellow Grant (P. E., 2020.0196) and a Scholar Grant (L. H. KAW 2016.0358). P.O.Å.P. also acknowledges the Swedish Foundation for Strategic Research (SSF) through project funding (EM16-0004) and the Research Infrastructure Fellow RIF 14-0074. Y. B. Li acknowledges the support from the China Postdoctoral Science Foundation (grant No. 2020M680082).

**Conflict of Interest**

The authors declare no conflict of interest.

**Data Availability**

The data that support the findings of this study are available from the corresponding author upon reasonable request.

# Supporting Information

Close-to-Room-Temperature Ferromagnetic Behavior of Single-Atom-Thick 2D Iron in Nanolaminated Ternary MAX Phases

*Youbing Li, Jinghua Liang, Haoming Ding, Jun Lu, Xulin Mu, Pengfei Yan, Xiao Zhang, Ke Chen, Mian Li, Per O. Å. Persson, Lars Hultman, Per Eklund, Shiyu Du, Hongxin Yang,\* Zhifang Chai, Qing Huang\**

Dr. Y. B. Li, H. M. Ding, X. Zhang, Dr. K. Chen, Dr. M. Li, Prof. S. Y. Du, Prof. Z. F. Chai, Prof. Q. Huang
Engineering Laboratory of Advanced Energy Materials
Ningbo Institute of Industrial Technology
Chinese Academy of Sciences
Ningbo, Zhejiang 315201, China
E-mail: huangqing@nimte.ac.cn
Dr. Y. B. Li, H. M. Ding, X. Zhang, Dr. K. Chen, Dr. M. Li, Prof. S. Y. Du, Prof. Z. F. Chai, Prof. Q. Huang
Qianwan Institute of CNiTECH
Zhongchuangyi Road, Hangzhou bay District
Ningbo, Zhejiang, 315336, China
Dr. J. H. Liang, Prof. H. X. Yang
Center of Materials Science and Optoelectronics Engineering
University of Chinese Academy of Sciences
E-mail: yanghongxin@nimte.ac.cn
Beijing 100049, China
Prof. J. Lu, Prof. P. O. Å. Persson, Prof. L. Hultman, Prof. P. Eklund
Department of Physics, Chemistry, and Biology (IFM)
Linköping University
Linköping 58183, Sweden
X. L. Mu, Prof. P. F. Yan
Beijing Key Laboratory of Microstructure and Properties of Solids
Beijing University of Technology
Beijing 100124, China
Y. B. Li and J. H. Liang contributed equally to this work

**Section 1 Experimental and Characterization**

**Materials**

Elemental powders of titanium nitride (300 mesh, 99.5% purity), titanium (300 mesh, 99.5% purity), tantalum (300 mesh, 99.5% purity), aluminum (300 mesh, 99.5% purity), niobium (300 mesh, 99.5 % purity), and niobium carbide (300 mesh, 99.5% purity) were commercially obtained from Target Research Center of General Research Institute for Nonferrous Metals, Beijing, China. Graphite (300 mesh, 99.5% purity), anhydrous ferrous chloride ($FeCl_2$, 98 wt.%), sodium chloride (NaCl, 98 wt.%), potassium chloride (KCl, 98 wt.%), (hydrochloric acid (HCl, 36.5%) and absolute ethanol ($C_2H_6O$, 98 wt.%) were commercially obtained from Aladdin Chemical Reagent, China.

**Preparation of precursor MAX phases of $Ti_2AlN$, $Nb_2AlC$ and $Ta_2AlC$**

TiN/Ti/Al/NaCl/KCl powder mixture with a mole ratio of 1: 1: 1.1: 8: 8 was used as the starting material and synthesized via a molten salt method. The mixture powders were fully mixed and placed in an alumina crucible. Then, the alumina crucible was put into a tube furnace and heated to 1100 °C at a rate of 5 °C/min for 7 h in a flowing Ar atmosphere. When the sample was cooled to room temperature, inorganic salts were removed by deionized water for 4-5 times. Finally, the sample was dried at 50 °C for 12h.

NbC/Al/Nb mixture powders with molar ratio of 1:1.1:1 were weighed and ball-milled in an agate jar for 48 h in order to synthesized $Nb_2AlC$ MAX phases. Then, the mixture of powders was uniaxially pressed in a graphite die under 10 MPa. Then the green compact was heated to 1600 °C for 30 min with a heating rate of 50 °C/min under the protection of argon. After the sample was cooled to room temperature at 20 °C/min, the obtained block was crushed and ground. Finally, the sample was passed through a 400 mesh sieve.

Ta/Al/C mixture powders with molar ratio of 2:1.1:1 were weighed and ball-milled in an agate jar for 24 h in order to synthesized $Ta_2AlC$ MAX phases. Then, the mixture of powders was uniaxially pressed in a graphite die under 10 MPa. Then the green compact was heated to 1550 °C with a heating rate of 50 °C/min under the protection of argon, and held 20 min under a pressure of 30 MPa. After the sample was cooled to room temperature at 20 °C/min, the obtained block was crushed and ground. Finally, the sample was passed through a 400 mesh sieve.

**Preparation of $M_2AX$ (A= Fe) phases**

The precursor MAX phases of $Ta_2AlC$, $Ti_2AlN$, and $Nb_2AlC$, and, and inorganic salts of $FeCl_2$ are in stoichiometric molar ratios of 2:3, respectively (experimental datils shown in Table S1). Then the mixed powders with NaCl/KCl were putted into an aluminum oxide boat and heated to reaction temperature during 7 h with a heating rate of 10 °C/min under an argon atmosphere. After the end of reaction, the extra salts remove by deionized water, and Fe metal was removed by hydrochloric acid about 2 h. Finally, the target products were obtained via filtering, washing, and drying at 50 °C in vacuum.

Table S1. The reaction conditions for synthesizing the novel MAX phases.

| MAX phases | Composition of raw materials (mole ratio) | T (°C) |
|---|---|---|
| $Ta_2FeC$ | $Ta_2AlC:FeCl_2:NaCl:KCl = 1:1.5:10:10$ | 650 |
| $Nb_2FeC$ | $Nb_2AlC:FeCl_2:NaCl:KCl = 1:1.5:10:10$ | 650 |
| $Ti_2FeN$ | $Ti_2AlN:FeCl_2:NaCl:KCl = 1:1.5:10:10$ | 600 |

**Characterization**

The phase composition of the samples was analyzed by X-ray diffraction (XRD, D8 Advance, Bruker AXS, Germany) with Cu $K_a$ radiation. The Rietveld method was employed to refine the crystal parameters and phase composition using thesoftware of TOPAS-Academic

v6. The microstructure and chemical composition were observed by scanning electron microscopy (SEM, QUANTA 250 FEG, FEI, USA) coupled with an energy-dispersive spectrometer (EDS). Structural and chemical analysis was carried out by a transmission electron microscope (TEM, FEI Tecnai G2 F20S-TWIN) with EDS at 300 kV and high-resolution STEM high angle annular dark field (HRSTEM-HAADF) imaging and STEM affiliated energy dispersive X-ray spectroscopy (EDS) within Link¨oping's double Cs corrected FEI Titan3 60-300 microscope operated at 300 kV.

The magnetic properties were measured on a Quantum Design superconducting quantum interference device magnetometer (SQUID). The powders were immersed in hydrochloric acid (HCl, 98 wt.%) for 2 h to remove the Fe metal. After washing, the powders were cold-pressed into round disks (diameter 10 mm) under a pressure of 20 MPa. A rectangular block of about 2 x 3 mm was cut the round disk, pasted on a quartz rod with tape and put into the SQUID instrument. Measurements were made in fields of 1000 Oe in the temperature range 2-400 K after cooling in zero applied field (ZFC) and in the measuring field (field-cooled, FC). The magnetic hysteresis loops measured in the fields of -10 kOe - 10 kOe at the temperature 2 K, 100 K, 200 K, 250 K, 300 K, and 400 K, respectively.

# Section 2 Characterization of M$_2$AX phases (A = Fe)

Table S2. Atomic positions in Ta$_2$FeC determined from the Rietveld refinement.

| Site | Element | x | y | z | Symmetry | Wyckoff symbol |
|---|---|---|---|---|---|---|
| M1 | Ta | 1/3 | 2/3 | 0.5948 | 3m | 4f |
| Al | Fe | 1/3 | 2/3 | 0.25 | -6m2 | 2d |
| C1 | C | 0 | 0 | 0 | -3m | 2a |

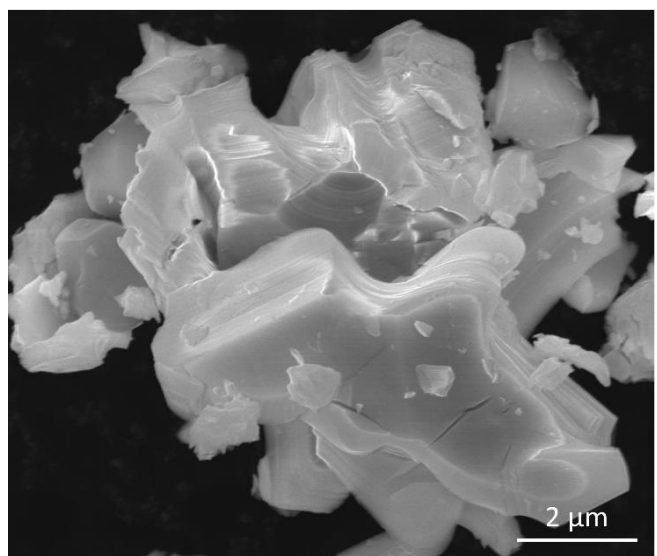

**Figure S1.** SEM image of Ta$_2$AlC MAX phase.

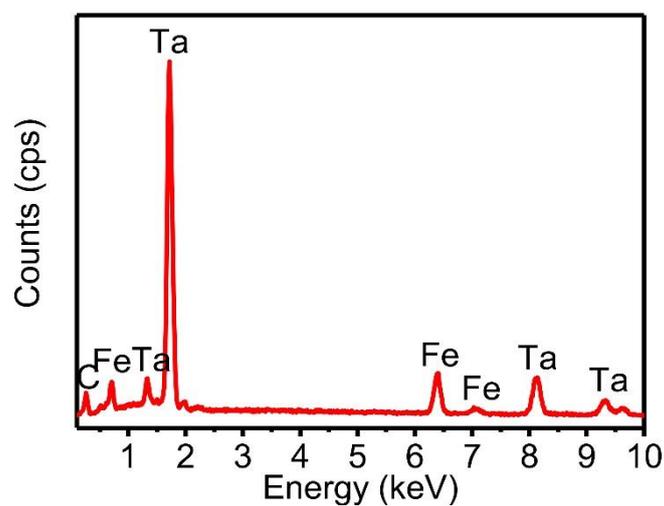

**Figure S2.** Energy-dispersive spectroscopy (EDS) spectrum of Ta$_2$FeC.

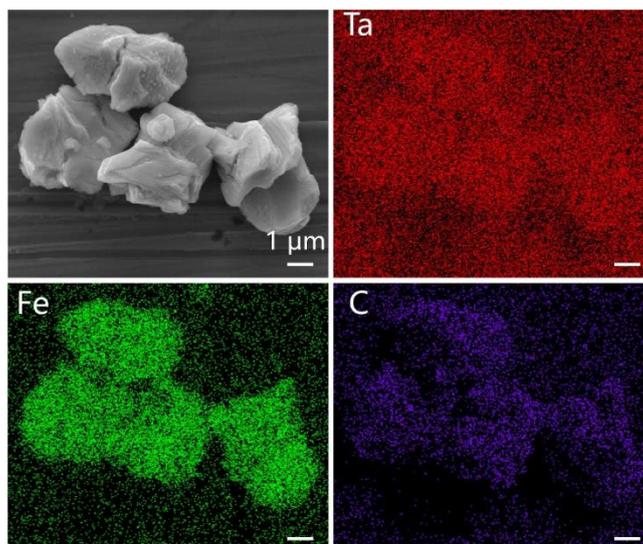

**Figure S3.** Elemental mapping showing the uniform distribution of Ta, Fe, and C elements in Ta$_2$FeC particles.

**Table S3.** Atomic positions in Ti$_2$FeN determined from the Rietveld refinement.

| Site | Element | x | y | z | Symmetry | Wyckoff symbol |
|---|---|---|---|---|---|---|
| M1 | Ti | 1/3 | 2/3 | 0.0848 | 3m | 4f |
| Al | Fe | 1/3 | 2/3 | 0.75 | -6m2 | 2d |
| C1 | N | 0 | 0 | 0 | -3m | 2a |

**Table S4.** Atomic positions in Nb$_2$FeC determined from the Rietveld refinement.

| Site | Element | x | y | z | Symmetry | Wyckoff symbol |
|---|---|---|---|---|---|---|
| M1 | Nb | 1/3 | 2/3 | 0.5957 | 3m | 4f |
| Al | Fe | 1/3 | 2/3 | 0.25 | -6m2 | 2d |
| C1 | C | 0 | 0 | 0 | -3m | 2a |

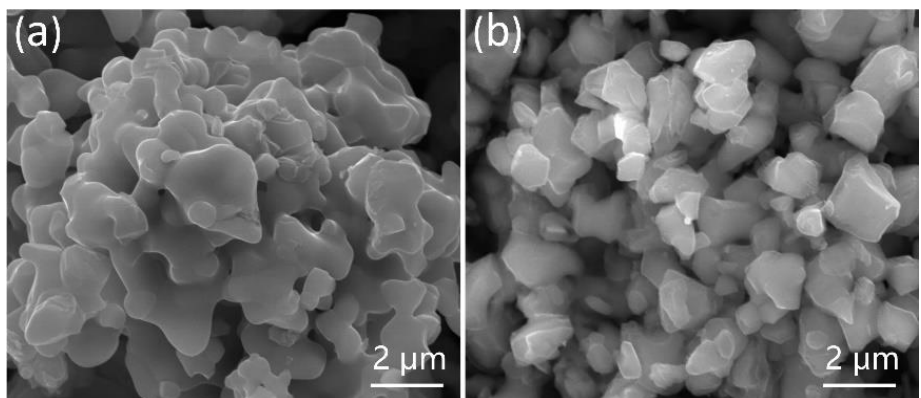

**Figure S4.** SEM images of Ti$_2$AlN (a) and Ti$_2$FeN (b) MAX phases.

**Figure S5.** SEM images of Nb$_2$AlC (a) and Nb$_2$FeC (b) MAX phases

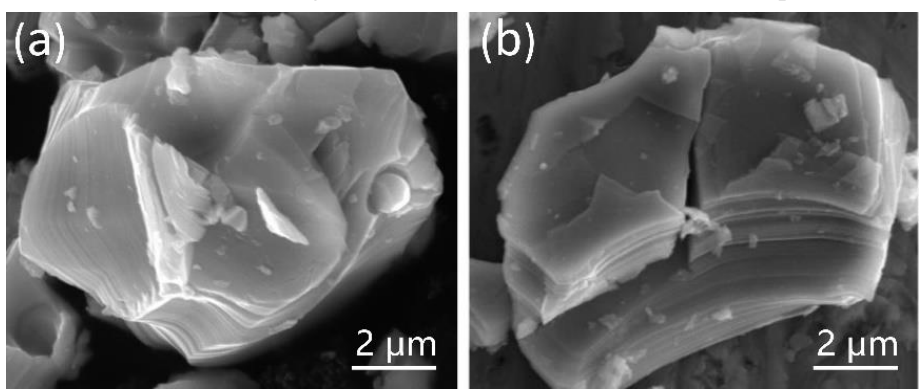

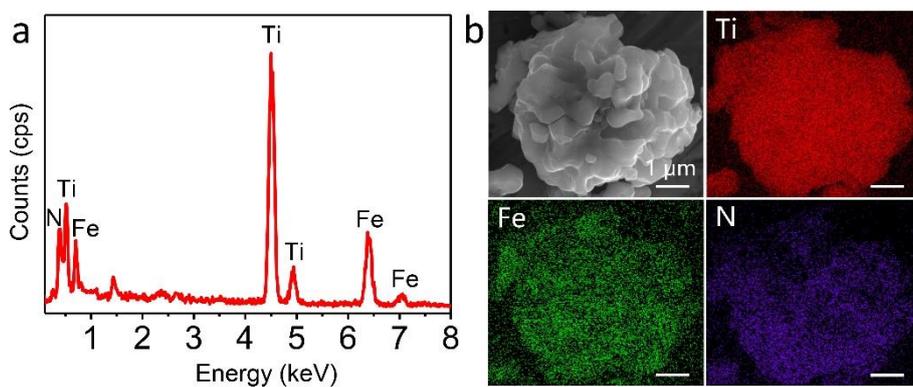

**Figure S6.** a) Energy-dispersive spectroscopy (EDS) spectrum of Ti$_2$FeN. b) Elemental mapping showing the uniform distribution of Ti, Fe, and N elements on Ti$_2$FeN particles.

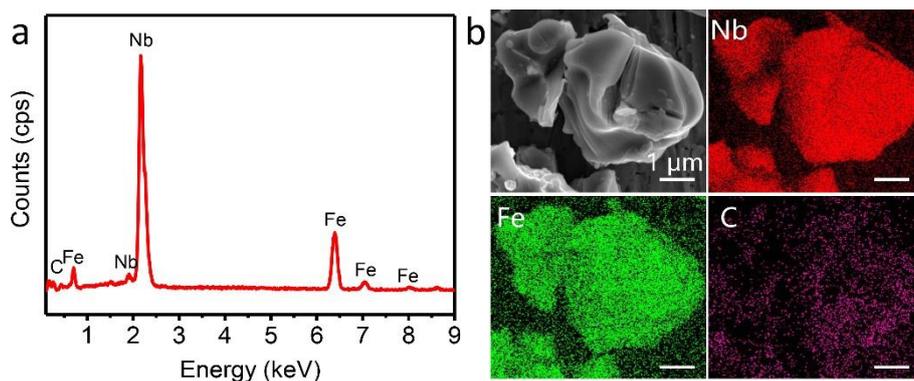

**Figure S7.** a) Energy-dispersive spectroscopy (EDS) spectrum of Nb$_2$FeC. b) Elemental mapping showing the uniform distribution of Nb, Fe, and C elements on Nb$_2$FeC particles.

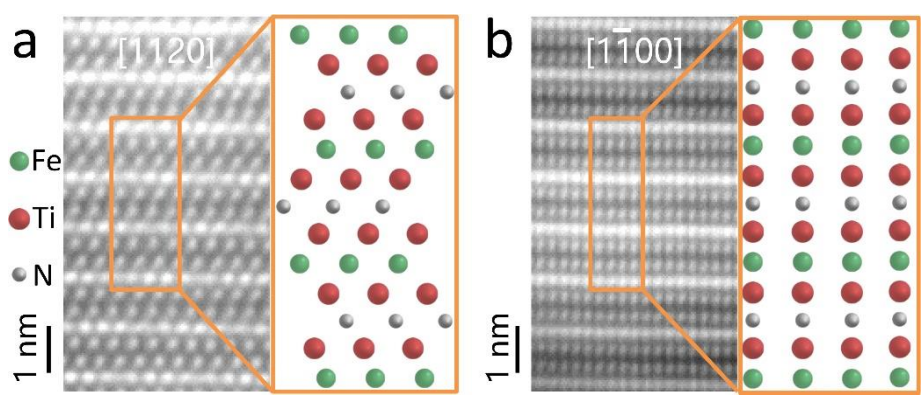

**Figure S8.** High-resolution (HR)-STEM images of Ti$_2$FeN showing atomic positions along (a) [11$\bar{2}$0] and (b) [1$\bar{1}$00] directions, respectively.

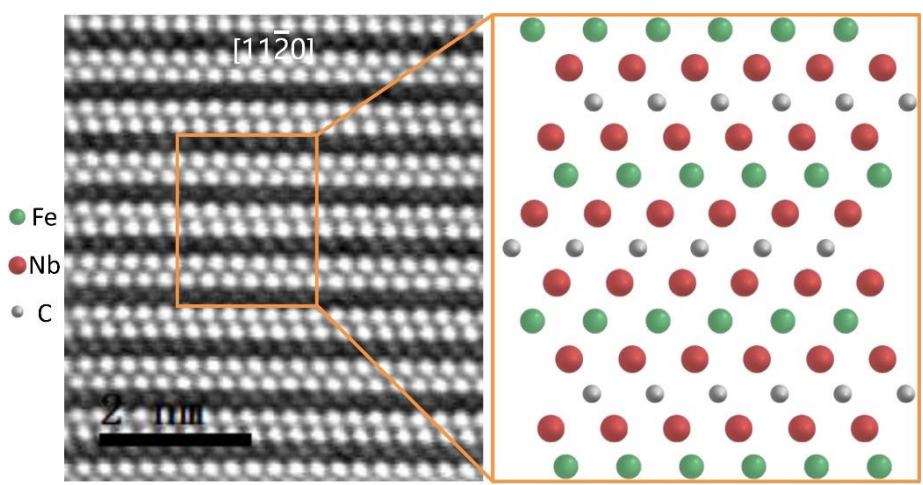

**Figure S9.** High-resolution (HR)-STEM image of Nb$_2$FeC showing atomic positions along [11$\bar{2}$0] direction.

## Section 3 Magnetic Properties

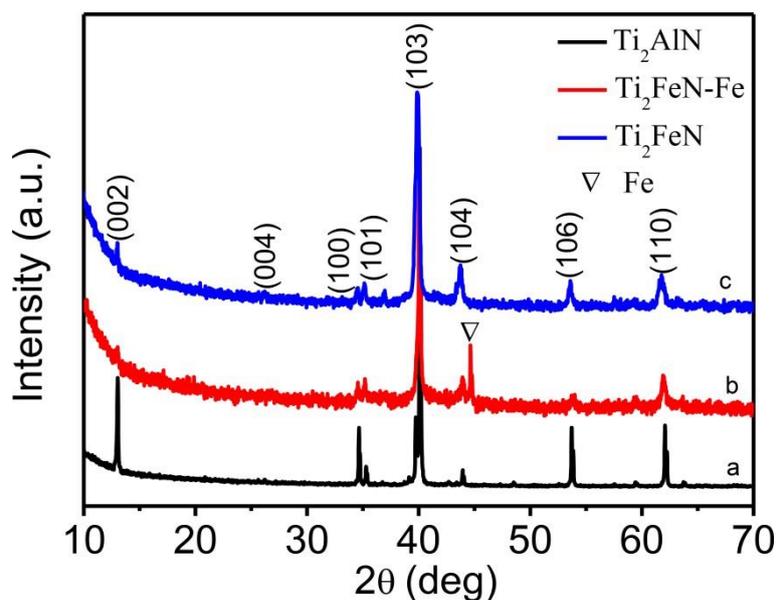

**Figure S10.** Phase compositions of magnetic MAX phase Ti$_2$FeN. a) XRD pattern of Ti$_2$AlN. b) After reaction of Ti$_2$AlN and FeCl$_2$, which is found that there is a clear Fe element. c) XRD patter of after wasded by HCl for 2 h, Fe diffraction peak was disappeared.

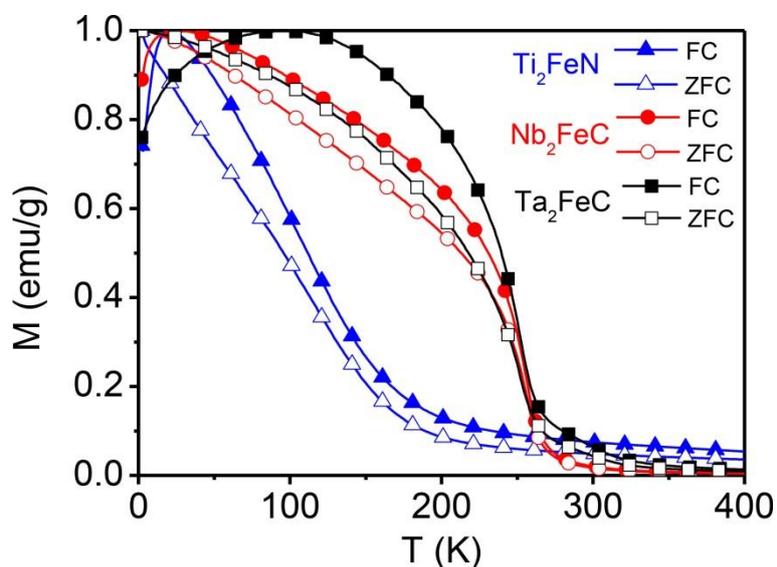

**Figure S11.** Zero-field-cooling (filled circles) and field-cooling (open circles) curves for the Ti$_2$FeN (blue lines), Nb$_2$FeC (red lines) and Ta$_2$FeC (black lines) at 1000 Oe with temperature in the range of 2-400 K.

**Table S5.** The coercive force ($H_c$), residual magnetization ($M_r$) and maximum saturation magnetization ($M_s$) of the Ta$_2$FeC at different temperature under ZFC in the magnetic field range of -1T to 1T.

| T (K) | $H_c$ (Oe) | $M_r$ (emu/g) | $M_s$ (emu/g) |
|---|---|---|---|
| 2 | 685.58 | 4.86 | 9.63 |
| 100 | 389.5 | 3.83 | 8.84 |
| 200 | 166.5 | 1.93 | 6.05 |
| 250 | 60.05 | 0.43 | 3.48 |
| 300 | 23.1 | 0.019 | 1.18 |
| 400 | 73.5 | 0.009 | 0.32 |

**Table S6.** The coercive force ($H_c$), residual magnetization ($M_r$) and maximum saturation magnetization ($M_s$) of the Ti$_2$FeN at different temperature under ZFC in the magnetic field range of -1T to 1T.

| T (K) | $H_c$ (Oe) | $M_r$ (emu/g) | $M_s$ (emu/g) |
|---|---|---|---|
| 2 | 625.19 | 1.671 | 5.93 |
| 20 | 234.5 | 0.975 | 4.96 |
| 50 | 146.5 | 0.645 | 4.44 |
| 100 | 68.5 | 0.259 | 3.59 |
| 150 | 11.7 | 0.013 | 2.26 |
| 200 | 11.6 | 0.007 | 1.01 |
| 300 | 11.7 | 0.006 | 0.39 |
| 400 | 29.5 | 0.008 | 0.24 |

**Table S7.** The coercive force ($H_c$), residual magnetization ($M_r$) and maximum saturation magnetization (M$_s$) of the Nb$_2$FeC at different temperature under ZFC in the magnetic field range of -1T to 1T.

| T (K) | $H_c$ (Oe) | $M_r$ (emu/g) | $M_s$ (emu/g) |
|---|---|---|---|
| 2 | 360.57 | 5.40 | 19.69 |
| 100 | 140.84 | 2.37 | 16.47 |
| 200 | 43.05 | 0.62 | 10.69 |
| 250 | 1.63 | 0.16 | 6.08 |
| 300 | 9.20 | 0.0031 | 1.49 |
| 400 | 23.86 | 0.0015 | 0.40 |

**Table S8.** The calculated total energy difference between the AFM1, AFM2 and FM state, the intralayer ($J_1$) and interlayer ($J_2$) exchange interaction parameters, the determined magnetic ground state, magnetic moments of Fe atom $\mu_{Fe}$, the saturation magnetization ($M_s$), Curie temperature $T_c$ as well as the experimental values of $\mu_{Fe}$, $M_s$ and $T_c$.

| Phase | $E_{AFM1} - E_{FM}$ (meV) | $E_{AFM2} - E_{FM}$ (meV) | $J_1$ (meV) | $J_2$ (meV) | Ground State | $\mu_{Fe}$ ($\mu_B$) | $M_s$ (emu/g) | $T_c$ (K) | $\mu_{Fe}$ ($\mu_B$) [Exp.] | $M_s$ (emu/g) [Exp.] | $T_c$ (K) [Exp.] |
|---|---|---|---|---|---|---|---|---|---|---|---|
| Ta$_2$FeC | 196.78 | 45.88 | 11.34 | 1.91 | FM | 1.16 | 13.65 | 240 | 0.74 | 9.63 | 291 |
| Ta$_2$FeN | 129.88 | -15.86 | 8.45 | -0.66 | AFM2 | 1.18 | -- | -- | -- | -- | -- |
| Nb$_2$FeC | 168.17 | 40.58 | 9.67 | 1.69 | FM | 1.29 | 25.86 | 210 | 0.89 | 19.69 | 281 |
| Nb$_2$FeN | -264.92 | 16.29 | -16.90 | 0.68 | AFM1 | 1.05 | -- | -- | -- | -- | -- |
| Ti$_2$FeC | 229.49 | -94.35 | 5.90 | -3.93 | AFM2 | 1.61 | -- | -- | -- | -- | -- |
| Ti$_2$FeN | 101.39 | 20.75 | 5.90 | 0.86 | FM | 0.94 | 22.51 | 130 | 0.18 | 5.93 | 208 |

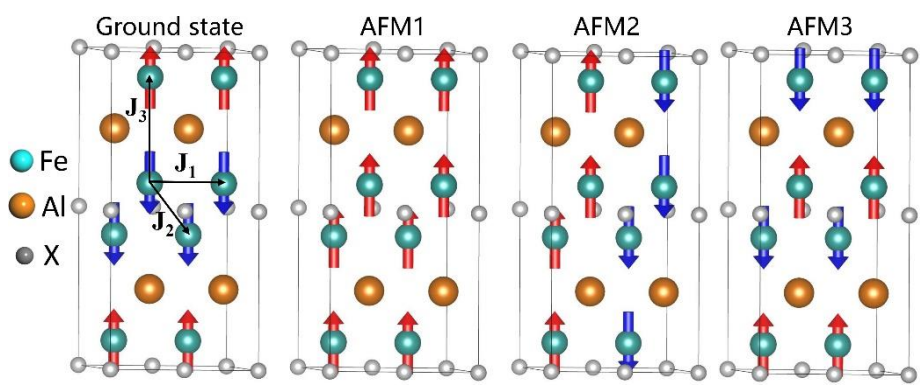

**Figure S12.** Considered intralayer (AFM1), interlayer (AFM2) and interlayer of two M$_2$X (AFM3) antiferromagnetic states for Fe$_2$AlC/Fe$_2$AlN MAX phases. The ground state of Fe$_2$AlC/Fe$_2$AlN is found to be antiferromagnetic.